\documentclass[useAMS,usenatbib]{mn2e}
\usepackage{epsfig}
\usepackage{deluxetable} 

\title[SBS 0846$+$513: a new $\gamma$-ray emitting Narrow-Line Seyfert 1 galaxy]{SBS 0846$+$513: a new $\gamma$-ray emitting Narrow-Line Seyfert 1 galaxy}
\author[F. D'Ammando, M. Orienti, J. Finke, et al.]{F. D'Ammando$^{1,2,3}$\thanks{E-mail: filippo.dammando@fisica.unipg.it}, M. Orienti$^{2,4}$, J. Finke$^{5}$, C. M. Raiteri$^{6}$, E. Angelakis$^{7}$, \newauthor L. Fuhrmann$^{7}$, M. Giroletti$^{2}$, T. Hovatta$^{8}$, W. Max-Moerbeck$^{8}$,  J. S. Perkins$^{9,10}$,  \newauthor A. C. S. Readhead$^{8}$, J. L. Richards$^{11}$, {\L.} Stawarz$^{12,13}$, D. Donato$^{9,10}$\\
$^{1}$Dip. di Fisica, Universit\'a degli Studi di Perugia, Via A. Pascoli, I-060123 Perugia, Italy \\
$^{2}$INAF - Istituto di Radioastronomia, Via Gobetti 101, I-40129 Bologna, Italy\\
$^{3}$CIFS, Viale Settimio Severo 63, I-10133 Torino, Italy \\
$^{4}$Dip. di Astronomia, Universit\'a di Bologna, Via Ranzani 1, I-40127 Bologna, Italy \\ 
$^{5}$U.S. Naval Research Laboratory, Code 7653, 4555 Overlook Ave. SW, Washington, DC 20375-5352, USA \\
$^{6}$INAF - Osservatorio Astronomico di Torino, Via Osservatorio 20, I-10025 Pino Torinese (TO), Italy \\
$^{7}$Max-Planck-Institute f\"ur Radioastronomie, Auf dem H\"ugel 69, D-53121 Bonn, Germany \\
$^{8}$Cahill Center for Astronomy and Astrophysics, California Institute of Technology 1200 E. California Blvd., Pasadena, CA 91125, USA \\
$^{9}$University of Maryland, Baltimore Country, 1000 Hilltop Circle, Baltimore, MD 20742, USA\\ 
$^{10}$Center for Research and Exploration in Space Science and Technology and NASA Goddard Space Flight Center, Greenbelt, MD 20771, USA\\
$^{11}$Department of Physics, Purdue University, 525 Northwestern Avenue, West Lafayette, IN 47907, USA\\
$^{12}$Institute of Space and Astronautical Science, JAXA, 3-1-1 Yoshinodai, Sagamihara, Kanagawa, 229-8510, Japan \\
$^{13}$Astronomical Observatory, Jagiellonian University, ul. Orla 171, Krak\'ow 30-244, Poland}
\begin{document}

\date{Accepted. Received; in original form}

\maketitle

\label{firstpage}

\begin{abstract}
We report {\em Fermi}-LAT observations of the radio-loud AGN SBS 0846$+$513 (z=0.5835), optically classified as a Narrow-Line Seyfert 1
galaxy, together with new and archival radio-to-X-ray data. The source was not active at $\gamma$-ray energies during the first two years of
{\em Fermi} operation. A significant increase in activity was observed during
2010 October--2011 August. In particular a strong $\gamma$-ray
flare was observed in 2011 June reaching an isotropic $\gamma$-ray luminosity (0.1--300 GeV) of 1.0$\times$10$^{48}$ erg s$^{-1}$, comparable to
that of the brightest flat spectrum radio quasars, and showing spectral evolution in $\gamma$ rays. An apparent superluminal velocity
of (8.2$\pm$1.5)$c$ in the jet was inferred from 2011-2012 VLBA images, suggesting the presence of a highly relativistic jet.

\noindent Both the power released by this object during the flaring activity and the apparent superluminal velocity are strong indications
of the presence of a relativistic jet as powerful as those of blazars. In addition, variability and spectral properties in radio and
$\gamma$-ray bands indicate blazar-like behaviour, suggesting that, except for
some distinct optical characteristics, SBS 0846$+$513 could be considered as a
young blazar at the low end of the blazar's black hole mass distribution.

\end{abstract}

\begin{keywords}
galaxies: active -- galaxies: nuclei -- galaxies: Seyfert -- galaxies:
individual (SBS 0846$+$513) -- gamma rays
\end{keywords}

\section{Introduction}

Active Galactic Nuclei (AGN) are the most luminous persistent sources
of high-energy radiation in the Universe. However, only a small
percentage of AGN are radio-loud, and this characteristic is commonly
ascribed to the presence of relativistic jets, roughly perpendicular
to the accretion disk. Accretion of gas on to the supermassive black
hole (SMBH) is thought to power these collimated jets, even if the
nature of the coupling between the accretion disc and the jet is still
among the outstanding open questions in high-energy astrophysics
\citep[e.g.,][]{blandford00,meier03}. Certainly relativistic jets are
the most extreme example of the power that can be generated by a SMBH
in the centre of an AGN, with apparent bolometric luminosities up to
10$^{49-50}$ erg s$^{-1}$ \citep[e.g.,][]{ackermann10,bonnoli11}, and
a large fraction of the power emitted in $\gamma$ rays.

Before the launch of the {\em Fermi} satellite only two classes of AGN
were known to generate these structures and thus to emit up to the
$\gamma$-ray energy range: blazars and radio galaxies, both hosted in
giant elliptical galaxies \citep{blandford78}. The first two years of
observations by the Large Area Telescope (LAT) on board {\em Fermi}
confirmed that these two are the most numerous classes of identified
sources in the extragalactic $\gamma$-ray sky \citep[]{abdo10a,
nolan12}, but the discovery of variable $\gamma$-ray emission from 4
radio-loud Narrow-Line Seyfert 1 galaxies (NLS1s) revealed the
presence of a possible emerging third class of AGN with relativistic
jets \citep[]{abdo09a, abdo09b, abdo09c}. In addition, {\em Fermi}-LAT
observations gave us the opportunity to study in more detail
particular subclasses of the established types of $\gamma$-ray
emitting AGN \citep[e.g.~Broad Line Radio Galaxies,][]{abdo10b,
kataoka11}. On the contrary, no radio-quiet Seyfert galaxies were
detected in $\gamma$ rays until now \citep{ackermann12}.

NLS1 are a class of AGN identified by~\citet{osterbrock85} and
characterized by the following optical properties: narrow permitted
lines (FWHM (H$\beta$) $<$ 2000 km s$^{-1}$) emitted from the broad
line region, [OIII]/H$\beta$ $<$ 3 \citep[a criterion introduced by][]{goodrich89}, and a bump due to Fe II \citep[see
e.g.][for a review]{pogge00}. They also exhibit strong X-ray
variability, steep X-ray spectra \citep[photon indices $\Gamma_{\rm X} > 2$;][]{boller96, grupe10}, and substantial soft X-ray excess \citep{boller96}. These
characteristics point to systems with smaller masses of the central black hole \citep[10$^6$-10$^8$ M$_\odot$,][]{yuan08} than in blazars and radio galaxies and higher accretion rates \citep[close to or above the Eddington limit,][]{yuan08}. NLS1 are generally radio-quiet ($R <10$, radio-loudness $R$ being defined as ratio of rest-frame 1.4\,GHz and 4400\,\AA\, flux densities), with only a small fraction classified as radio-loud \citep[$<$ 7$\%$,][]{komossa06}, even more sparse ($\sim$2.5$\%$) are very  radio-loud NLS1s ($R > 100$), while generally 10$\%$-15$\%$ of quasars are radio-loud and very radio-loud. In the past, several authors
investigated the peculiarities of radio-loud NLS1s with
non-simultaneous radio to X-ray data, suggesting similarities with
the young stage of quasars or different types of blazars
\citep[]{komossa06,yuan08,foschini09}. The strong and variable radio
emission, and the flat radio spectrum together with variability
studies suggested the presence in some radio-loud NLS1s of a
relativistic jet, confirmed in some objects by the {\em Fermi}-LAT
detection in $\gamma$ rays \citep{abdo09c}. This finding poses intriguing questions about the nature of these objects, the onset of production of relativistic jets, and the cosmological evolution of radio-loud AGN. The impact of  peculiar characteristics of the central engines in radio-loud NLS1s, which seem quite different of those of blazars and manifest in their peculiar optical characteristics, on the $\gamma$-ray production mechanisms is currently under debate.

In June 2011 high $\gamma$-ray activity from SBS 0846$+$513 was
observed by {\em Fermi}-LAT. Preliminary  results were reported in \citet{donato11}. That was an important
confirmation of the detection of a new $\gamma$-ray NLS1 after the
claim by~\citet{foschini11}\footnote{The analysis presented
in~\citet{foschini11} was performed with the P6\_V3 IRFs, starting
from the first Fermi-LAT catalogue (1FGL) as a reference for the
background sources, and over the period August 2008--February 2011.}.

 First identified with a BL Lac object at redshift z = 1.86 during a
 bright state by \citet{arp79}, SBS 0846$+$513 is a high redshift
 object positioned very close to a low redshift galaxy at the end of a
 chain of five galaxies. \citet{arp79} noted also a small nebulosity
 close to SBS 0846$+$513 that could be a normal red galaxy observed by
 chance near the object due to projection effects. The Sloan Digital
 Sky Survey (SDSS) spectrum of the source reported in \citet{zhou05}
 is typical of NLS1: FWHM (H$\beta$) = (1710 $\pm$ 184) km s$^{-1}$,
 [OIII]/H$\beta$ $\simeq$ 0.32, and a strong Fe II bump. Similar
 values were obtained by \citet{yuan08} analysing the same SDSS
 spectrum: FWHM (H$\beta$) = (1810 $\pm$ 191) km s$^{-1}$ and
 [OIII]/H$\beta$ $\simeq$ 0.31.  Moreover, the SDSS spectrum showed
 that its true redshift is z = 0.5835. \citet{nottale86} and others
 suggested that it was a gravitationally lensed quasar with
 variability due to gravitational amplification by a star in an
 intervening galaxy. However, with broader wavelength coverage and
 high resolution, no signs of any intervening galaxies have been found
 in the SDSS spectrum. The Hubble Space Telescope (HST) image,
 collected when SBS 0846$+$513 was in a faint state ($V\sim$19.7 mag)
 does not show significant resolved structure, thus no indication of
 the host galaxy \citep{maoz93}. High polarization ($>$ 10$\%$) was
 detected in optical and radio by \citet{moore81} and
 \citet{sitko84}. A remarkable optical variability was observed in the
 past with an amplitude of $\Delta V \sim 5$\ mag over 1 year and
 $\Delta V \sim 4$ mag over 1 month \citep{arp79}. The high
 polarization, very high brightness temperature \citep[$T_{b}>
 10^{13}$\ K;][]{zhou05}, and the large-amplitude variability observed
 in optical were clues for the presence of an at least mildly
 relativistic jet, now confirmed by the LAT detection in $\gamma$ rays
 and superluminal motion revealed by our new radio observations.

In this paper we present the detection of SBS 0846$+$513 in $\gamma$
rays and discuss its characteristics by means of the available
radio-to-$\gamma$-ray data. Preliminary results about the high
resolution VLBA observations of the source were presented in
\citet{orienti12}. The paper is organized as follows: in Section 2 we
report the LAT data analysis and results. In Section 3 we report the
result of the {\em Swift} observations performed in August--September
2011. Radio data collected by Medicina, OVRO, Effelsberg, VLBA, and
VLA are presented in Section 4 and discussed in Section 5 and 6. In
Section 7 we present the spectral energy distribution (SED) modeling, while
discussion and concluding remarks are presented in Section 8.

Throughout the paper, a $\Lambda$--CDM cosmology with $H_0$ = 71 km
s$^{-1}$ Mpc$^{-1}$, $\Omega_{\Lambda} = 0.73$, and $\Omega_{m} =
0.27$ is adopted. The corresponding luminosity distance at $z =
0.5835$ is d$_L = 3406$\ Mpc, and 1 arcsec corresponds to a projected
distance of 6.584 kpc.

\section{{\em Fermi}-LAT Data: Selection and Analysis}
\label{FermiData}

The {\em Fermi}-LAT is a $\gamma$-ray telescope operating from $20$\,MeV to
$>300$\,GeV. The instrument is an array of $4 \times 4$ identical
towers, each one consisting of a tracker (where the photons are
pair-converted) and a calorimeter (where the energies of the
pair-converted photons are measured). The entire instrument is covered
with an anticoincidence detector to reject the charged-particle
background. The LAT has a large peak effective area ($\sim$ $8000$\,cm$^2$ for
$1$\,GeV photons), an energy resolution typically $\sim$10\%,
and a field of view of about $ 2.4$ \,sr with an angular
resolution ($68\%$ containment angle) better than 1\degr\ for
energies above $1$\,GeV. Further details about the LAT
are given by \citet{atwood09}.

\begin{figure}
\centering
\includegraphics[width=7.5cm]{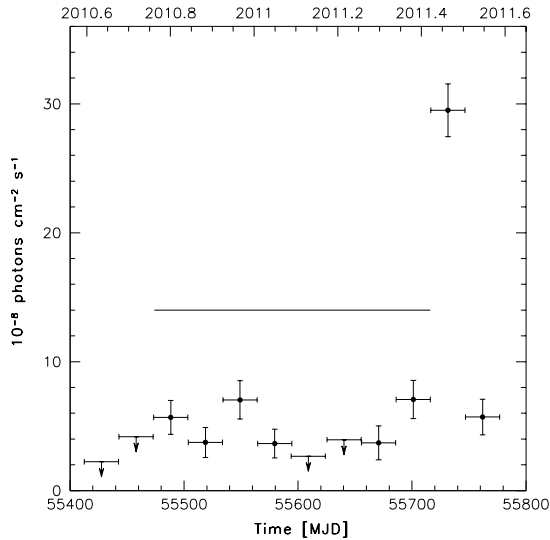}
\caption{Monthly integrated flux (E $>$ 100 MeV) light curve of SBS\,0846+513
  obtained from 2010 August 4 to 2011 August 4. Arrows refer to 2-$\sigma$
  upper limits on the source flux. Upper limits are computed when TS $<$
  10. The horizontal line indicates the period considered for the SED.}
\label{Fig1}
\end{figure}

The LAT data reported in this paper were collected over the first 40
months of {\em Fermi} operation, from 2008 August 4 (MJD 54682) to
2011 December 4 (MJD 55899). During this time the LAT instrument
operated almost entirely in survey mode. The analysis was performed
with the \texttt{ScienceTools} software package version v9r23p1. The
LAT data were extracted within a $15^{\circ}$ Region of Interest (RoI) centred at the radio location of SBS 0846$+$513. Only events
belonging to the ``Source'' class were used. In addition, a cut on the
zenith angle ($< 100^{\circ}$) was also applied to reduce
contamination from the Earth limb $\gamma$ rays, which are produced by
cosmic rays interacting with the upper atmosphere. The spectral
analysis (from which we derived spectral fits and photon fluxes) were
performed with the post-launch instrument response functions (IRFs)
\texttt{P7SOURCE\_V6} using an unbinned maximum likelihood method
implemented in the Science tool \texttt{gtlike}.

The background model used to extract the $\gamma$-ray signal includes
a Galactic diffuse emission component and an isotropic component. The
model that we adopted for the Galactic component is given by the file gal\_2yearp7v6\_v0.fits, and the isotropic component, which is the sum of
the extragalactic diffuse emission and the residual charged particle
background, is parametrized by the file iso\_p7v6source.txt\footnote{http://fermi.gsfc.nasa.gov/ssc/data/access/lat/Background\\Models.html}.
The normalizations of both components in the background model were allowed to vary freely during the spectral point fitting. 

We examine the significance of the $\gamma$-ray signal from the
sources by means of the Test Statistics (TS) based on the likelihood
ratio test. The Test Statistic TS = 2$\Delta$log(likelihood) between
models with and without the source is a measure of the probability of
having a $\gamma$-ray source at the localization specified, which compares
models whose parameters have been adjusted to maximise the likelihood of the
data given the model \citep{mattox96}. The source model used in \texttt{gtlike}
includes all the point sources from the second {\em Fermi}-LAT catalogue
\citep[2FGL;][]{nolan12} that fall within $20^{\circ}$ from the source. The spectra of these sources were parametrized by power-law functions,
except for 2FGL J0920.9$+$4441, and 2FGL J0957.7$+$5522, for which we used a log-parabola in their spectral modeling as in the 2FGL
catalogue. We removed from the model the sources having test statistic TS $<$
25 and/or fluxes below 1.0$\times$10$^{-8}$ photons cm$^{-1}$ s$^{-1}$ over 40
months and repeated the fit. Thus a final fitting procedure has been performed
with the sources within 10$^{\circ}$ from
SBS 0846$+$513 included with the normalization factors and the photon indices left as free parameters. 
For the sources located between 10$^{\circ}$ and 15$^{\circ}$ we kept the
normalization and the photon index fixed to the values obtained in the
previous fitting procedure. The RoI model includes also sources falling between 15$^{\circ}$ and
20$^{\circ}$ from the target source, which can contribute to the total counts observed in the RoI
due to the energy depended size of the point spread function of the instrument. For these
additional sources, normalizations and indices were also fixed to the values of the 2FGL
catalog.

SBS 0846$+$513 was not in the 1FGL or 2FGL catalogues, indicating that
the source was not detected with TS $>$ 25 in either one year or two
years of {\em Fermi} observations
\citep[]{abdo10a,nolan12}. Integrating over the first two years of
{\em Fermi} operation the fit yielded a TS = 14. The 2-$\sigma$ upper
limit over the first 2 years of LAT data is 8.5$\times$10$^{-9}$
photons cm$^{-2}$ s$^{-1}$ in the 0.1--300 GeV energy
range\footnote{To obtain a convergence we performed the fit
considering only the sources within $10^{\circ}$ from SBS 0846+513 and
E $>$ 300 MeV and extrapolated the flux down to 100 MeV.}, assuming
the photon index is $\Gamma=2.3$, which is the average value found for
low-synchrotron-peaked blazars in the Second LAT AGN Catalog (2LAC)
\citep{ackermann12}. On the contrary, the fit with a power-law model
to the data integrated over the third year of {\em Fermi} operation
(2010 August 4 -- 2011 August 4; MJD 55412--55777) in the 0.1--300 GeV
energy range results in a TS = 653, with an integrated average flux of
(6.7 $\pm$ 0.5) $\times$10$^{-8}$ photons cm$^{-2}$ s$^{-1}$ and a
photon index $\Gamma$ = 2.23 $\pm$ 0.05.

Finally, over the period 2011 August 4--December 4 (MJD 55777--55899)
only a TS = 8 was obtained, with a 2-$\sigma$ upper limit of
1.9$\times$10$^{-8}$ photons cm$^{-2}$ s$^{-1}$ in the 0.1--300 GeV
energy range, assuming the photon index is $\Gamma=2.23$. During
August--December 2011 the source was not detected with TS $>$ 10 using
a time bin of 1 month, indicating a significant variability in the
$\gamma$-ray activity of the source on month time-scales. For the rest
of the paper we focused on the $\gamma$-ray data collected over 2010
August 4--2011 August 4.

The $\gamma$-ray
point source localization by means of the \texttt{gtfindsrc} tool applied to
the $\gamma$ rays extracted during the third year of observation results in R.A. =
132.48 deg, Dec. = $+$51.13 deg (J2000), with a 95$\%$ error circle radius of
0.06$^{\circ}$, at an angular separation of 0.01$^{\circ}$ from the radio position
of SBS 0846$+$513 (R.A. = 132.492 deg, Dec. = $+$51.141 deg, J2000). This implies a
strict spatial association with the radio coordinates of the NLS1 SBS
0846$+$513. The proposed association could be questioned to some extent
  due to the presence of a low redshift star-forming galaxy almost on the line
  of sight toward SBS 0846$+$513. And indeed, several members of the Local
  Group and also some nearby starburst galaxies have been recently detected by {\em
    Fermi}-LAT and thus established as sources of high energy photons. Their
  $\gamma$-ray luminosities range however between 10$^{40}$ and
  $\sim$3 $\times$10$^{43}$ erg s$^{-1}$, and no variability in their
  $\gamma$-ray emission was neither observed nor expected \citep{abdo10e}. The
  very large $\gamma$-ray luminosity of the LAT object discussed here and its
  flaring activity therefore rule out the foreground galaxy as the source of
  the detected $\gamma$ rays.

\begin{table*}
\caption{Unbinned likelihood spectral fit results.}
\begin{tabular}{lcc|ccc|ccc}
 \hline \hline
                    &  \multicolumn{1}{c}{PL}                  &   \multicolumn{2}{c}{LP}                                    &   \multicolumn{2}{c}{BPL}\\
Time Period (MJD)& $\Gamma$ &$\alpha$ &$\beta$  &$\Gamma_1$ &$\Gamma_2$ \\
\hline
55412--55777 & 2.23 $\pm$ 0.05 & 1.93 $\pm$ 0.12 & 0.13 $\pm$ 0.05 & 2.00
$\pm$ 0.09 & 2.70 $\pm$ 0.18 \\       
55717--55747 & 1.98 $\pm$ 0.05 & 1.23 $\pm$ 0.19 & 0.30 $\pm$ 0.07& 1.51 $\pm$ 0.12 & 2.81 $\pm$ 0.21 \\
\hline
  \end{tabular}
\label{LAT}
\end{table*}

In order to test for curvature in the $\gamma$-ray spectrum of SBS
0846$+$513 two alternative spectral models with respect to the
power-law (PL) were used: the log parabola (LP), $dN/dE \propto$
$E/E_{0}^{-\alpha-\beta \, \log(E/E_b)}$ \citep[]{landau86,
massaro04}, and the broken power law (BPL). In the case of the log
parabola the parameter $\alpha$ is the spectral slope at the energy
$E_0$ and the parameter $\beta$ measures the curvature around the
peak. We fixed the reference energy $E_0$ to 300 MeV.  We used a
likelihood ratio test to check the PL model (null hypothesis) against
the LP model (alternative hypothesis). These values may be compared,
following \citet{nolan12}, by defining the curvature Test Statistic
TS$_{\rm curve}$=2(log L$_{\rm LP}$-log L$_{\rm PL}$)=10 corresponding
to $\sim$3.3-$\sigma$ difference. This value is below the threshold of
TS$_{\rm curve}$=16 applied in \citet{nolan12} for defining a
significant curvature, thus suggesting only a hint of spectral
curvature in the average $\gamma$-ray spectrum of SBS 0846$+$513.  We
tested also the broken power-law model, with photon indices $\Gamma_1$
below and $\Gamma_2$ above the break energy $E_{\rm break}$. We fixed
$E_{\rm break}$ at 1.4 GeV. This value was estimated studying the
profile of the Likelihood function, fixing $E_{\rm break}$ at
different values between 100 MeV and 3 GeV with a step of 50 MeV and
finding the minimum log(likelihood) value.  Also for this spectral model
we obtain TS$_{\rm curve}$=2(log L$_{\rm BPL}$-log L$_{\rm PL}$)=10,
therefore we adopt the power-law for the following LAT analysis. The
fit results are reported in Table~\ref{LAT}.

\begin{figure}
\centering
\includegraphics[width=7.5cm]{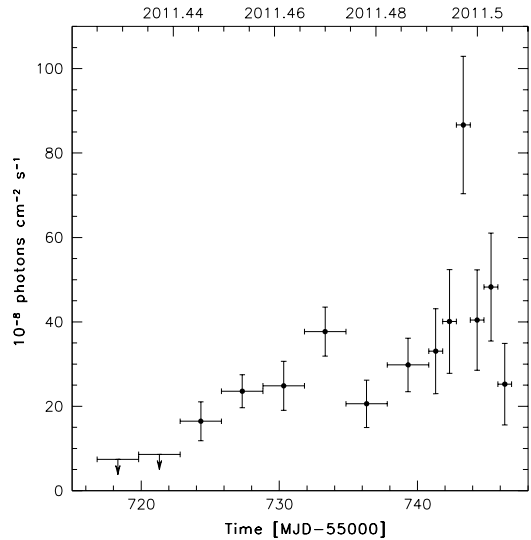}
\caption{Integrated flux (E $>$ 100 MeV) light curve of SBS\,0846+513
  obtained from 2011 June 4 to 2011 July 4 with 3-day or 1-day time bins. Arrows refer to 2-$\sigma$
  upper limits on the source flux. Upper limits are computed when TS $<$ 10.}
\label{Fig2}
\end{figure}

\begin{table*}
\caption{Log and fitting results of {\em Swift}/XRT observations of
  SBS 0846$+$513 using a power-law model with $N_{\rm H}$ fixed to Galactic
  absorption. $^{a}$Observed flux.}
\begin{center}
\begin{tabular}{cccc}
\hline \hline
\multicolumn{1}{c}{Observation} &
\multicolumn{1}{c}{Net Exposure Time} &
\multicolumn{1}{c}{Photon index} &
\multicolumn{1}{c}{Flux 0.3-10 keV$^{a}$} \\
\multicolumn{1}{c}{Date} &
\multicolumn{1}{c}{sec} &
\multicolumn{1}{c}{$\Gamma$} &
\multicolumn{1}{c}{$\times$10$^{-13}$ erg cm$^{-2}$ s$^{-1}$} \\
\hline
2011-08-03 & 6738 & $1.51 \pm 0.24$ & $9.0 \pm 2.2$ \\
2011-09-15 & 5427 & $1.43 \pm 0.28$ & $7.7 \pm 2.9$ \\
\hline
\end{tabular}
\end{center}
\label{XRT}
\end{table*}

Figure~\ref{Fig1} shows the $\gamma$-ray light curve of the third year of {\em Fermi}
observations built using 1-month time bins. For each time bin the photon index
of SBS 0846$+$513 and all sources within 15$^{\circ}$ from it was frozen to the value resulting from the likelihood
analysis over the entire year, while for the sources between 15$^{\circ}$ and 20$^{\circ}$ from SBS
0846$+$513 the photon index was frozen to the value reported in the 2FGL
catalogue. If TS $<$ 10 the value of the fluxes were replaced by the 2-$\sigma$ upper limits. The systematic uncertainty in
the flux is energy dependent: it amounts to $10\%$ at 100 MeV, decreasing to
$5\%$ at 560 MeV, and increasing to $10\%$ above 10 GeV. By means of the \texttt{gtsrcprob} tool we estimated that the
highest energy photon emitted by SBS 0846$+$513 (with probability $>$80\% of
being associated with the source) was observed on 2 May 2011 at distance of
0.34$^{\circ}$ from the source with an energy of 31 GeV. 

The source was not continuously detected over the entire third year and the flux remained below
10$^{-7}$ ph cm$^{-2}$ s$^{-1}$, except in the period 2011 June 4--July 4 when an increase in flux by a factor of $\sim$ 4 was observed. Considering the high activity of the source we extracted a spectrum over
that period, obtaining a photon index $\Gamma$ = 1.98 $\pm$ 0.05 and a flux of
(24.4 $\pm$ 2.1) $\times$10$^{-8}$
photons cm$^{-2}$ s$^{-1}$. 
A similar spectral evolution has been already observed in other flat spectrum
radio quasars (FSRQs) during a bright state \citep{abdo10d}. A light curve
focused on the period 2011 June 4--July 4, extracted with the photon index fixed to $\Gamma$ = 1.98, was
built with 3-day and 1-day time bins (Fig.~\ref{Fig2}). We used a 1-day time bin for
the period with higher statistics. The peak of the emission was observed
between June 30 19:43 UT and July 1 19:43 UT, with a flux of (87 $\pm$ 16) $\times$10$^{-8}$
photons cm$^{-2}$ s$^{-1}$ in the 0.1--300 GeV energy range. The peak
$\gamma$-ray flux is a factor of $\sim$13 higher with respect to the average
flux estimated over the entire third year of observation. The $\gamma$-ray
flare is characterized by a sharp increase in flux of a factor of $\sim$2.5 in
1 day and return to the previous flux level in 1 day. 

\noindent For taking into
consideration the possible influence of the choice of the T$_{0}$ (the start
time of the bin) used for
building the light curve on the determination of the $\gamma$-ray peak we shifted the T$_{0}$ of 12 hours back and forth and
recalculated the daily flux, obtaining (53 $\pm$ 15) $\times$10$^{-8}$
and (66 $\pm$ 15) $\times$10$^{-8}$ photons cm$^{-2}$ s$^{-1}$,
respectively.
Finally, during the month of high activity, replacing the power-law with a
log parabola or a broken power-law we obtain a TS$_{curve}$ = 49, indicating a
significant curvature. We noted that the curvature parameter $\beta$ in the
log-parabola spectral model increased as the flux increased (see Table~\ref{LAT}).

\section{Swift Data: Analysis and Results}
\label{SwiftData}

The {\em Swift} satellite \citep{gehrels04} performed two observations
of SBS 0846$+$513 on 2011 August 30 and September 15 for 6.8 ksec and
5.5 ksec, respectively. The observations were performed with all three onboard intruments: the X-ray Telescope (XRT; \citet{burrows05},
0.2--10.0 keV), the Ultraviolet Optical Telescope (UVOT;
\citet{roming05}, 170--600 nm) and the Burst Alert Telescope (BAT;
\citet{barthelmy05}, 15--150 keV).

The hard X-ray flux of this source is below the sensitivity  of the BAT
instrument for such short exposures and therefore the
data from this instrument will not be used. Moreover, the source was not present in the
{\em Swift} BAT 58-month hard X-ray catalogue \citep{baumgartner10} and the 54-month Palermo BAT catalogue \citep{cusumano10}.

The XRT data were processed with standard procedures (\texttt{xrtpipeline v0.12.6}), filtering, and screening criteria by using the \texttt{Heasoft} package
(v6.11). The data were collected in photon counting mode in both of the observations, and only XRT event grades 0--12 were selected. The source count rate
was low ($<$ 0.5 counts s$^{-1}$); thus pile-up correction was not required. Source events were extracted from a circular region with a radius of
20 pixels (1 pixel $\sim$ 2.36$"$), while background events were extracted from
a circular region with radius of 50 pixels away from the source
region. Ancillary response files were generated with \texttt{xrtmkarf}, and
account for different extraction regions, vignetting and PSF corrections. We
used the spectral redistribution matrices v013 in the Calibration database
maintained by HEASARC. 

Considering the low number of photons collected ($<$ 200 counts) the
spectra are rebinned with a minimum of 1 count per bin and the Cash
statistics \citep{cash79} are used. We fit the spectrum with an
absorbed power-law using the photoelectric absorption model
\texttt{tbabs} \citep{wilms00}, with a neutral hydrogen column fixed
to its Galactic value \citep[2.91$\times$10$^{20}$
cm$^{-2}$;][]{kalberla05}. The low photon statistics prevents us from fitting the X-ray data with a more complicated model than a simple
power-law. We noted that also in \citet{grupe10}, in which a large sample of
NLS1s observed by {\em Swift}/XRT was studied, the absorption column density
was fixed to the Galactic value and the authors found that in the majority of
NLS1s the spectrum can be fit sufficiently well by an absorbed single
power-law model with negligible intrinsic absorption.
The fit results are reported in
Table~\ref{XRT}.  The relatively hard X-ray spectrum with respect to
the other NLS1s \citep[e.g.][]{grupe10} could be due to the
contribution of inverse Compton radiation from a relativistic jet,
similarly to flat spectrum radio quasars. A photon index 1.5--1.8 was
observed in X-rays also for PMN J0948$+$0022, the first NLS1 detected
in $\gamma$ rays by {\em Fermi}-LAT \citep{abdo09b, foschini_etal11}.

In the past SBS 0846$+$513 was detected in X-rays only by ROSAT, with a 0.1-2.4
keV flux of 2.7$\times$10$^{-13}$ erg cm$^{-2}$ s$^{-1}$ and a photon index
$\Gamma$ = 1.77$^{+0.44}_{-0.60}$ \citep{yuan08}. As a comparison the
  flux observed on August 30 by {\em
  Swift}/XRT in the same energy range with $\Gamma$ fixed to 1.77 is $3.7 \pm
0.6$ $\times$10$^{-13}$ erg cm$^{-2}$ s$^{-1}$, $\sim$40$\%$ higher than the
ROSAT observations, adding evidence of the variability character of the source.

In 2011 August the UVOT instrument took 5 frames in the $w1$ band only,
while four sequences with all $v$, $b$, $u$, $w1$, $m2$, and $w2$
filters were acquired in 2011 September. These data were processed with v6.10 of the \texttt{Heasoft} package and the CALDB release
dated 2011 August 12. Source counts were extracted from a circular
region of 5\arcsec\ radius centred on the source, while background
counts were derived from a circular region of 10\arcsec\ radius in the
source neighbourhood.  By fitting the source spectrum of September 15
with a power-law, we calculated the effective wavelengths, count rate
to flux conversion factors ($\rm CF_\Lambda$), and Galactic
extinctions for the UVOT bands according to the procedure explained in
\citet{raiteri10,raiteri11}. The results are shown in
Table~\ref{caluvot}. To obtain the UVOT de-reddened fluxes reported in
Table~\ref{res_uvot} we multiplied the count rates for the $\rm
CF_\Lambda$ and corrected for the corresponding Galactic extinction
values $\rm A_\Lambda$.

We note that a {\em Swift}/UVOT observation performed on 2011 September
15 found SBS 0846$+$513 about 0.4 magnitude brighter in V-band with
respect to the HST observation in January 1992.

\begin{table}
\caption{Results of the UVOT calibration procedure: effective wavelengths
  $\lambda_{\rm eff}$, count rate to flux conversion factors $\rm CF_\Lambda$,
  and Galactic extinction calculated from the \citet{cardelli89} laws.}
\label{caluvot}      
\centering                          
\begin{tabular}{l c c c}       
\hline\hline                 
Filter & $\lambda_{\rm eff}$ & $\rm CF_\Lambda$ & $A_\Lambda$ \\    
       & \AA\ & $10^{-16} \rm \, erg \, cm^{-2} \, s^{-1} \, \AA^{-1}$ & mag \\
\hline                        
   $v$    & 5430 & 2.60 & 0.09 \\      
   $b$    & 4360 & 1.47 & 0.11 \\
   $u$    & 3476 & 1.65 & 0.14 \\
   $uvw1$ & 2621 & 4.40 & 0.19 \\
   $uvm2$ & 2257 & 8.36 & 0.24 \\ 
   $uvw2$ & 2087 & 5.97 & 0.23 \\ 
\hline                                  
\end{tabular}
\end{table}
  
\begin{table}
\caption{Results of the analysis of UVOT data for SBS 0846$+$513.}
\label{res_uvot}      
\centering                          
\begin{tabular}{l c c c}       
\hline\hline                 
MJD & Filter & Magnitude & Magnitude$\_$error \\    
\hline
55803.644 &  $uvw1$ & 19.57 & 0.07 \\                        
55819.497 &  $v$    & 19.10 & 0.30 \\      
55819.489 &  $b$    & 19.67 & 0.19 \\
55819.488 &  $u$    & 19.22 & 0.17 \\
55819.497 &  $uvw1$ & 19.41 & 0.15 \\
55819.501 &  $uvm2$ & 19.55 & 0.15 \\ 
55819.493 &  $uvw2$ & 19.55 & 0.10 \\ 
\hline                                  
\end{tabular}
\end{table}

\section{Radio Data: Analysis and Results}\label{RadioData}

New and archival radio data for SBS 0846$+$513 were collected from Medicina,
OVRO, the Effelsberg 100-m, VLA and VLBA. In the following subsections we
present the observations performed by these facilities.

\subsection{Medicina}

We observed SBS\,0846+513 with the Medicina radio telescope five times between
2011 August and November. 
We used the new Enhanced Single-dish Control System (ESCS) acquisition system, which provides enhanced
sensitivity and supports observations with the cross scan technique. All
observations were performed at both 5 and 8.4 GHz; the typical on source time
is 1.5 minutes and the flux density was calibrated with respect to 3C 286. Since
the signal to noise ratio in each scan across the source was low (typically
$\sim 3$), we performed a stacking analysis of the scans, which allowed us to
significantly improve the signal to noise ratio and the accuracy of the
measurement. We list the final values of the 5 and 8.4 GHz flux density in
Table~\ref{t.medicina}.

\begin{table}
\begin{center}
\caption{Results of the Medicina 32-m radio observations. \label{t.medicina}}
\begin{tabular}{lcc}
\hline \hline
Obs.\ date & $S_5$ & $S_{8.4}$ \\
 & (mJy) & (mJy) \\
\hline
2011-08-10 & $170 \pm 15$ & $235 \pm 20$ \\
2011-09-08 & $210 \pm 20$ & $185 \pm 20$ \\
2011-09-22 & $190 \pm 20$ & $180 \pm 20$ \\
2011-10-13 & $190 \pm 20$ & $280 \pm 30$ \\
2011-11-16 & $250 \pm 30$ & $275 \pm 25$ \\
\hline
\end{tabular}
\end{center}
\end{table}

\subsection{OVRO}

\begin{figure}
\centering
\includegraphics[width=7.5cm]{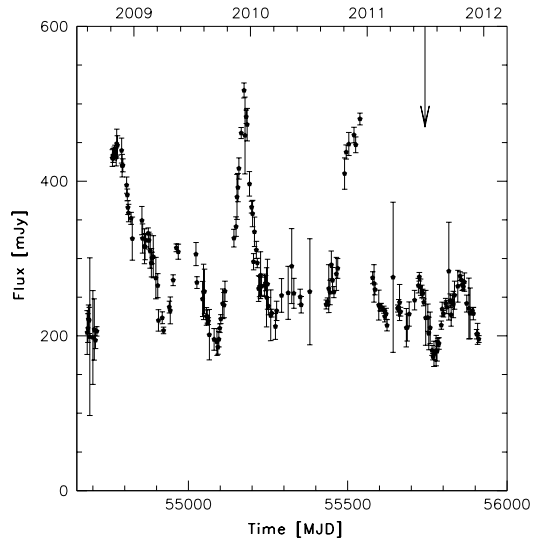}
\caption{15 GHz radio light curve for the period 2008 August 4--2011 December 12 from the OVRO telescope. The downward arrow indicates the time of the peak of the
  $\gamma$-ray activity observed by {\em Fermi}-LAT.}
\label{Fig3}
\end{figure}

As part of an ongoing blazar monitoring program, the Owens Valley
Radio Observatory (OVRO) 40~m radio telescope has observed SBS 0846$+$513
at 15~GHz regularly since the end of 2007 \citep{richards11}. This
monitoring program includes over 1500 known and likely $\gamma$-ray loud
blazars above declination $-20^{\circ}$. The sources in this
program are observed in total intensity twice per week with a 4~mJy
(minimum) and 3\% (typical) uncertainty. Observations are performed
with a dual-beam (each 2.5~arcmin FWHM) Dicke-switched system using
cold sky in the off-source beam as the reference. Additionally, the
source is switched between beams to reduce atmospheric variations. The
absolute flux density scale is calibrated using observations of
3C~286, adopting the flux density (3.44~Jy) from \citet{baars77}. 
This results in about a 5\% absolute scale uncertainty, which is not reflected in the plotted errors.

\subsection{Effelsberg 100-m}

The centimeter spectrum of SBS 0846+513 was observed with the Effelsberg 100-m
telescope on 2011 April 30 (MJD 55681.8) within the framework of a
{\em Fermi}-related monitoring program of $\gamma$-ray blazars \citep[F-GAMMA program;][]{fuhrmann07}. The measurements were conducted with the
secondary focus heterodyne receivers at 2.64, 8.35, 14.60, and 32.00 GHz. The
observations were performed quasi-simultaneously with cross-scans, that is,
slewing over the source position, in azimuth and elevation directions, with
adaptive numbers of sub-scans for reaching the desired sensitivity \citep[for details, see][]{fuhrmann08, angelakis08}. Pointing offset correction, gain correction, atmospheric opacity correction, and sensitivity correction have been applied to the data.

\subsection{VLA and VLBA data}

\begin{table*}
\caption{Log of the archival VLA observations and flux density.}\label{vla}
\begin{center}
\begin{tabular}{cccccc}
\hline \hline
Freq&Date&Code&Obs.time&Beam&Flux\\
GHz &    &    &min&arcsec$\times$arcsec&mJy\\
\hline
1.4&1986-04-10&AV127&9.16&1.40$\times$1.12&194$\pm$6\\
4.8&1986-04-10&AV127&9.16&0.44$\times$0.35&286$\pm$9\\
4.8&1996-01-05&BW021&1.66&1.39$\times$1.19&332$\pm$10\\
4.8&1996-12-30&BA018&3.33&0.43$\times$0.36&363$\pm$11\\
4.8&2009-05-09&CALSUR&0.66&2.12$\times$1.18&196$\pm$6\\
8.4&1995-08-13&AM484&0.5&0.24$\times$0.22&356$\pm$11\\
8.4&1995-09-02&AM484&0.5&0.32$\times$0.23&327$\pm$10\\
\hline
\end{tabular}
\end{center}
\end{table*}

To implement the information on the flux density variability available from
the single-dish observations and to study the source structure on kpc and pc
scales, we analysed archival VLA and VLBA data at different frequencies obtained between 1986 and 2011. To study the proper motion we also made use of 15-GHz VLBA data from the MOJAVE programme\footnote{The MOJAVE data 
archive is maintained at http://www.physics.purdue.edu/MOJAVE.}.\\
Usually SBS 0846$+$513 was observed with the VLA as a phase calibrator and the on-source 
observing time is generally quite short. Logs of VLA and VLBA observations are
reported in Tables~\ref{vla} and ~\ref{vlba}, respectively. \\
The data reduction of both VLA and VLBA data was performed following the standard procedures implemented in the National Radio Astronomy
Observatory (NRAO) \textsc{aips} package. In the case of VLA data, the accuracy of the
amplitude calibration was checked by means of either 3C\,48 or 3C\,286, and it
resulted to be within 3\%. For the VLBA datasets, a priori amplitude calibration was derived using measurements of the system temperature
and the antenna gains. Uncertainties of the amplitude calibrations were estimated to be within 5-10\%. For MOJAVE
data we imported the fully calibrated {\em uv} datasets \citep{lister09}. \\
The final VLA and VLBA images were produced after a number of phase-only
self-calibration iterations. The source parameters, like flux density and angular size, were derived on the radio image plane by means of the \textsc{aips} task
JMFIT which performs a Gaussian fit to the source and its subcomponents
(Table~\ref{vlba_results}). For the extended component visible at 15 GHz, we
determine the flux density by means of TVSTAT, which performs an aperture integration on a selected region on the image plane.

\begin{figure}
\includegraphics[width=7.5cm]{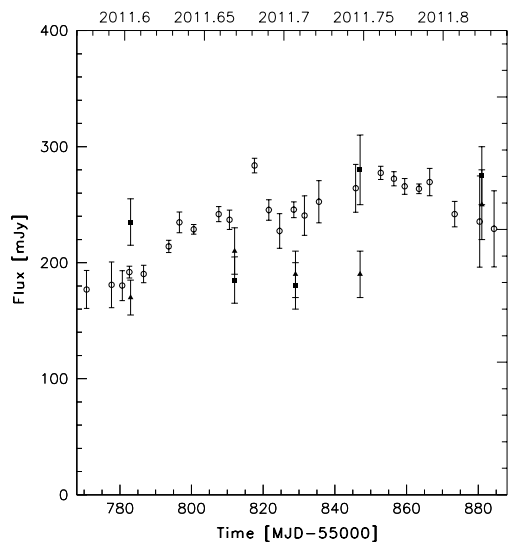}
\caption{Radio data of SBS\,0846+513 collected at 5.0 GHz (triangles) and 8.4
  GHz (squares) by Medicina, and 15 GHz (open circles) by OVRO during the
  period 2011 July 28 - November 19 (MJD 55770-55884).}
\label{Fig4}
\end{figure}

\begin{figure*}
\centering
\includegraphics[width=7.0cm]{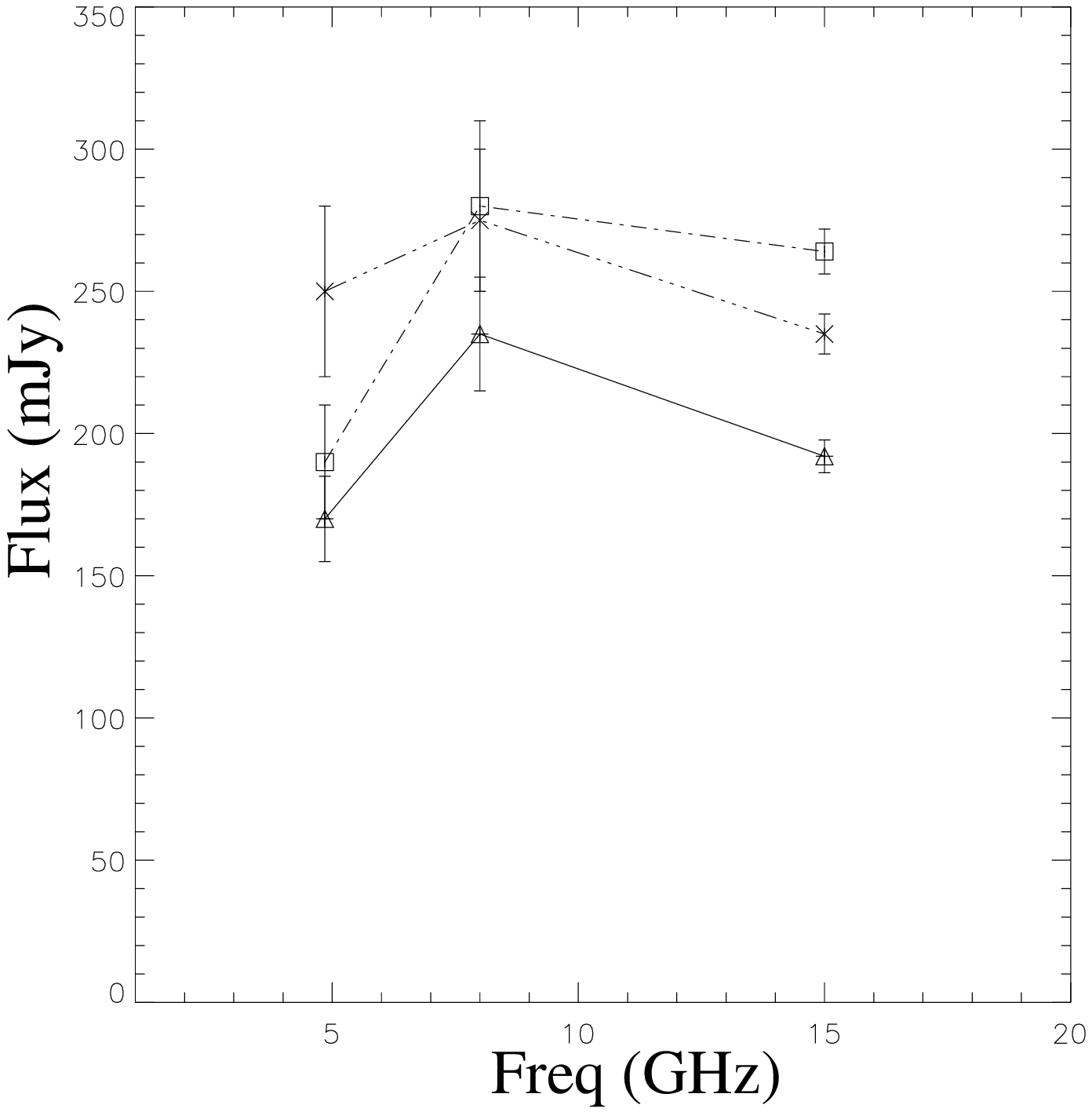}
\hspace{3mm}
\includegraphics[width=7.0cm]{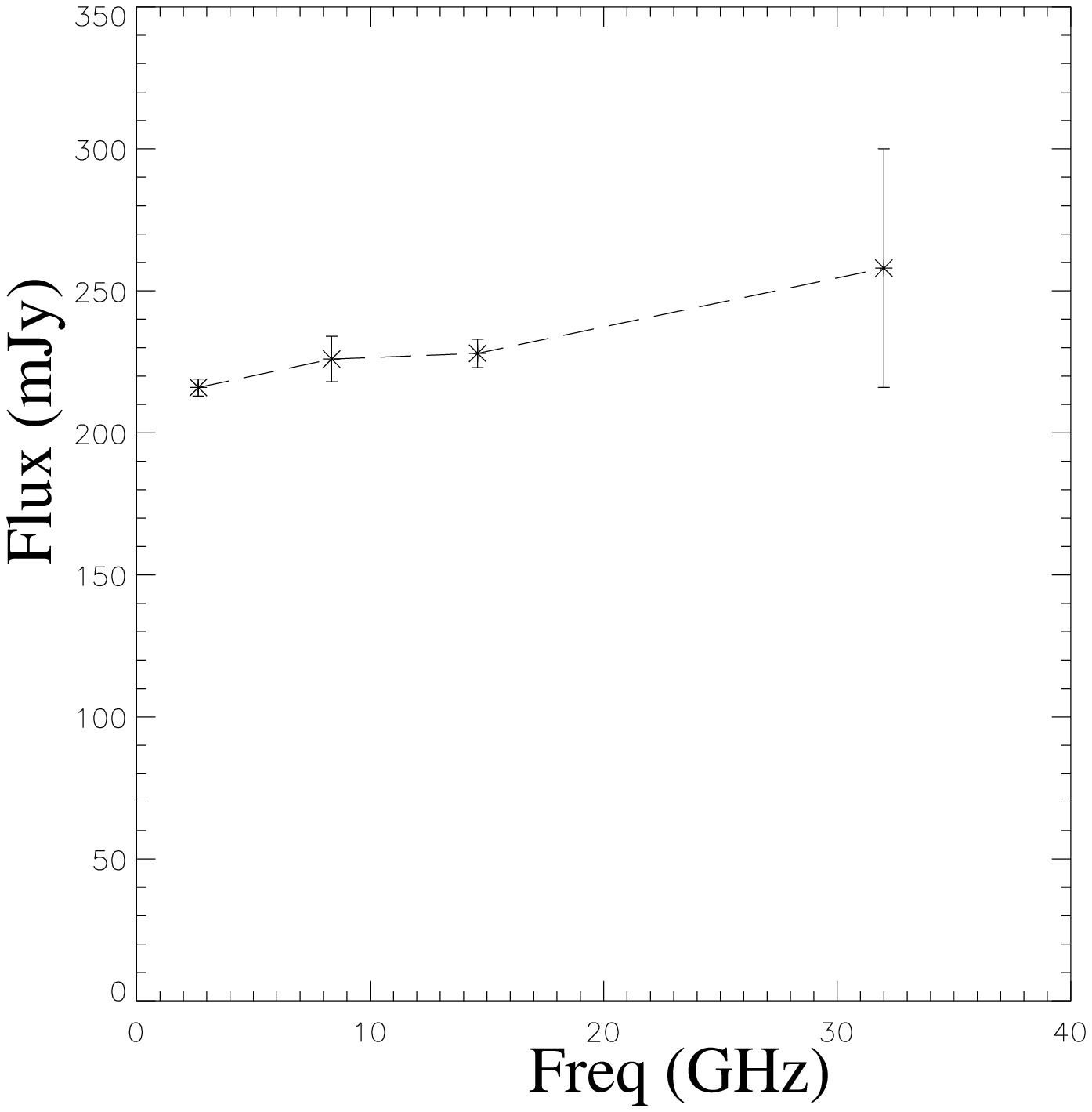}
\caption{{\em Left panel}: Radio spectra of SBS 0846+513. Observations
at 5 and 8.4 GHz are from the Medicina single-dish telescope, those at
15 GHz are from OVRO. Triangles, squares, and crosses refer to
observations performed around 2011 August 10, October 13, and November
16, respectively. {\em Right panel}: Radio spectrum of SBS 0846+513
obtained by the Effelsberg single-dish on 2011 April 30 from 2.64 to
32 GHz.}
\label{Fig5and6}
\end{figure*}

\section{Radio variability}

Multi-epoch studies of the radio emission of SBS\,0846+513 show significant flux density variability. In Fig.~\ref{Fig3} we report the OVRO
observations obtained since the launch of {\em Fermi}. From the analysis of the OVRO light curve at 15 GHz it is clear that high and low activity states are
interspersed. Interestingly, no $\gamma$-ray emission was detected during the
major radio flaring episodes that occurred before 2011. A maximum flux density
variation of a factor of $\sim$2.0 was observed over 3 months during the
period 2010 August--2011 August, while in the period 2008 August--2010 August a
variation of a factor of $\sim$2.8 was observed over a similar time interval.\\ 
Flux density variability is detected also at 4.8 GHz, where the flux density varies about 70\% during the time interval spanned by the 
VLA data (1986 to 2009). However, the poor time coverage of the VLA data does not allow us
to fully characterize the flux density behaviour on such a long time interval.\\
The detection of the $\gamma$-ray flare in 2011 June, triggered a
monitoring campaign with the Medicina telescope. In Fig.~\ref{Fig4}  we
report the Medicina flux density at 5 ({\em triangles}) and 8.4 GHz
({\em squares}) in addition to the OVRO 15-GHz data ({\em open circles})
collected between 2011 July 28 and November 19. From 
August 20 (MJD 55793.628), i.e.~about one month after the
$\gamma$-ray flare, the flux density at 15 GHz started to
increase reaching the maximum of about 280 mJy around October 18, and
then it started to decrease. At lower frequencies the flux density
variation is observed after a longer time, firstly at 8.4 GHz (October 13),
and then at 5 GHz (November 16). This time delay may be
explained by opacity effects, which are more severe at longer wavelengths. During this time interval also the shape of the radio spectrum
changes, as is clearly visible in Fig.~\ref{Fig5and6} (left panel). We must note that
Medicina and OVRO observations are not strictly simultaneous and to build the
radio spectrum we considered the OVRO observation that is the closest
in time to the Medicina observing epoch.
For comparison, in Fig.~\ref{Fig5and6} (right panel) we show the simultaneous radio spectrum between 2.64 and 32 GHz
obtained by Effelsberg on 2011 April 30, i.e.~before the high activity state detected in $\gamma$ rays.

\section{Radio morphology}\label{radiomorph}

\begin{figure}
\centering
\includegraphics[width=7.5cm]{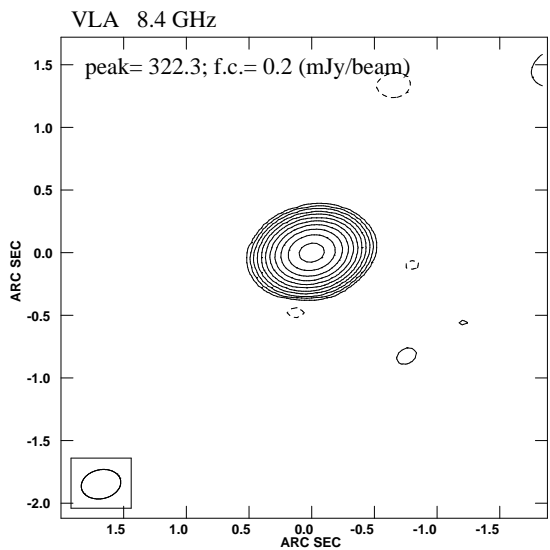}
\caption{VLA image at 8.4 GHz of SBS\,0846+513. On the image we provide the
peak flux density, in mJy/beam, and the first contour intensity (f.c.,
in mJy/beam) that corresponds to three times the noise measured on the
image plane. Contour levels increase by a factor of 2. The beam is
plotted on the bottom left corner of the image.}
\label{Fig7}
\end{figure}

\begin{figure}
\centering
\includegraphics[width=7.5cm]{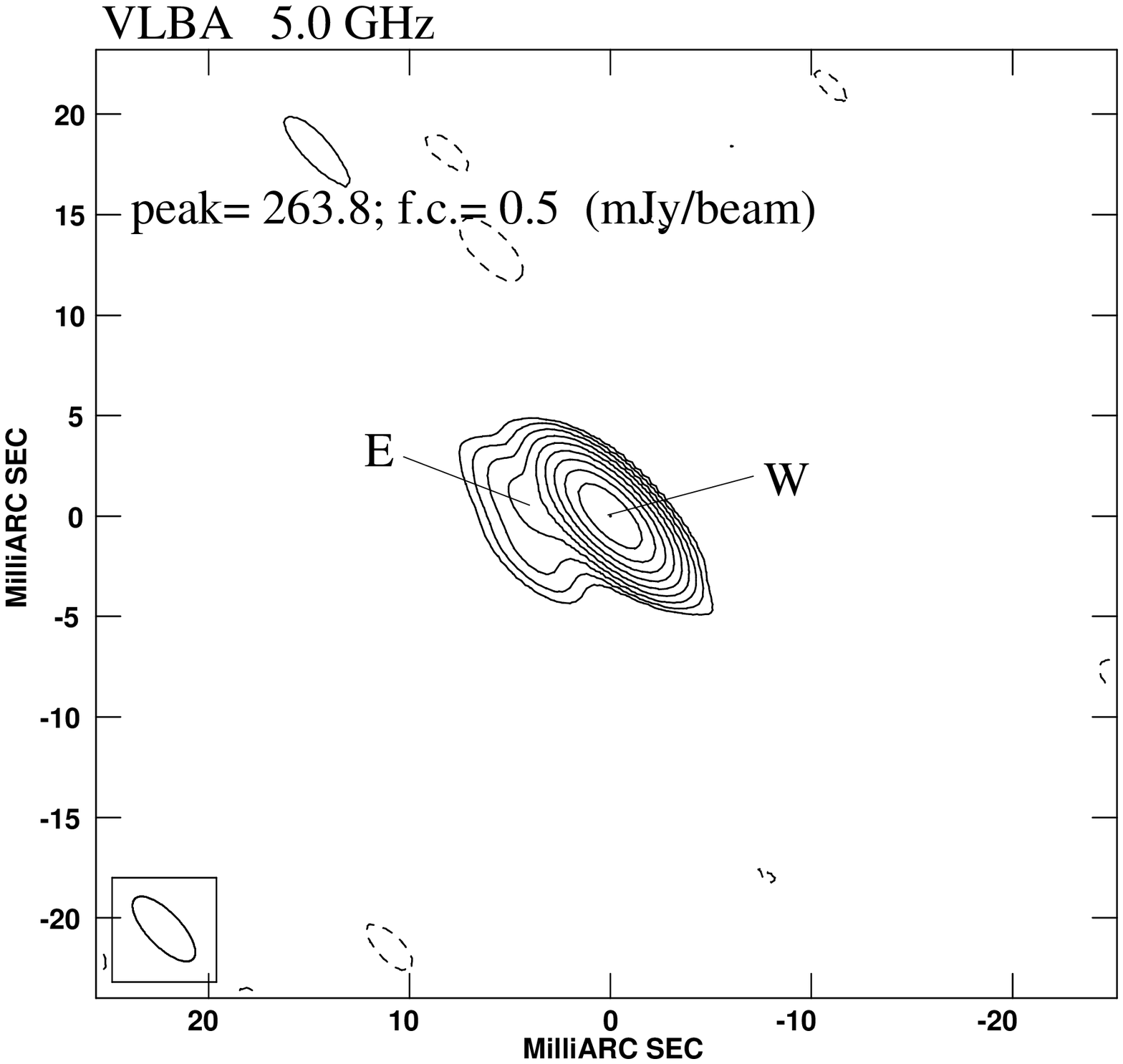}
\caption{VLBA image at 5.0 GHz of
SBS\,0846+513. On the image we provide the
peak flux density, in mJy/beam, and the first contour intensity (f.c.,
in mJy/beam) that corresponds to three times the noise measured on the
image plane. Contour levels increase by a factor of 2. The beam is
plotted on the bottom left corner of the image.}
\label{Fig8}
\end{figure}

\begin{figure}
\centering
\includegraphics[width=7.5cm]{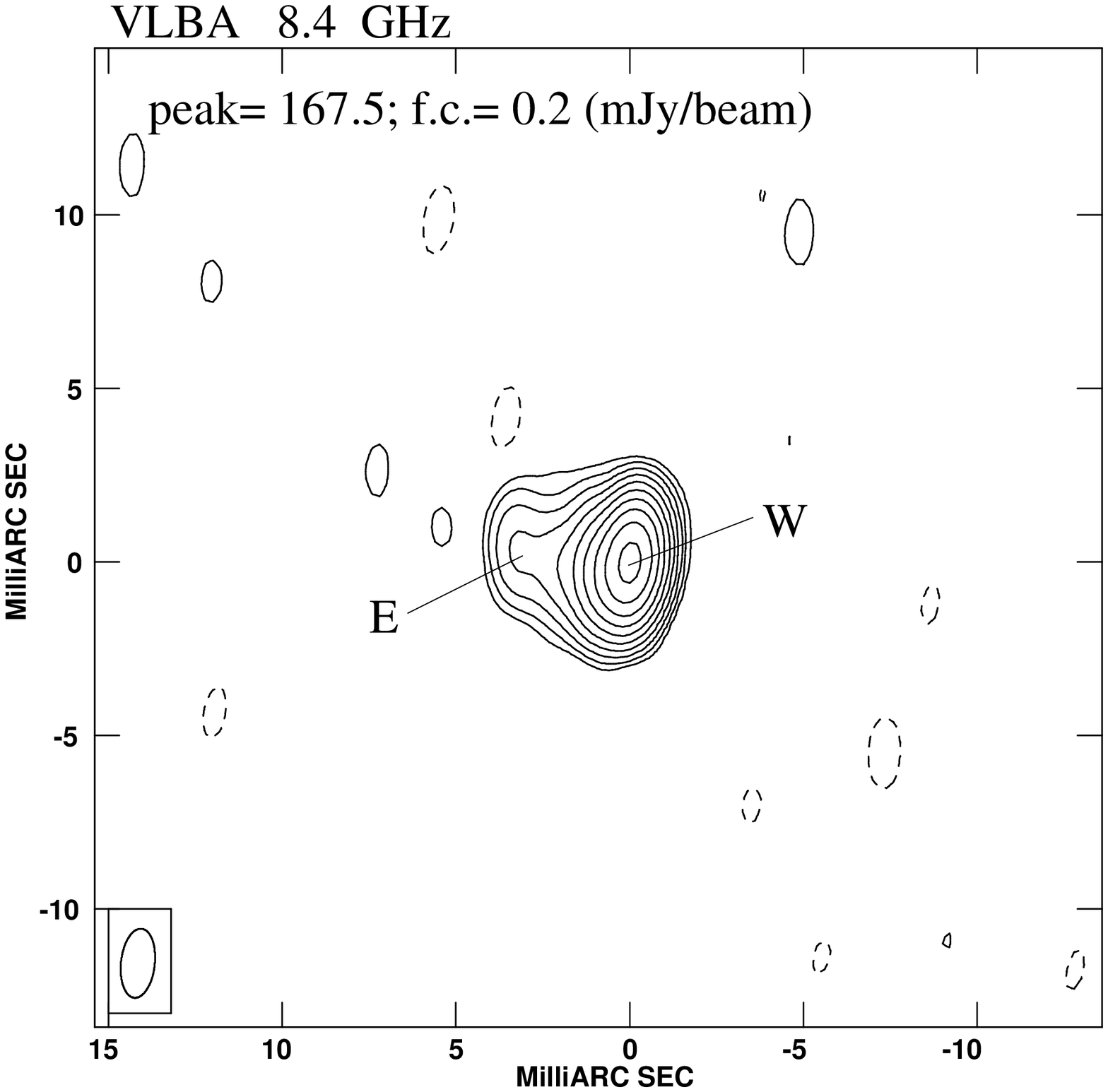}
\caption{VLBA image at 8.4 GHz ({\em bottom}) of
SBS\,0846+513. On the image we provide the
peak flux density, in mJy/beam, and the first contour intensity (f.c.,
in mJy/beam) that corresponds to three times the noise measured on the
image plane. Contour levels increase by a factor of 2. The beam is
plotted on the bottom left corner of the image.}
\label{Fig9}
\end{figure}

\begin{figure}
\centering
\includegraphics[width=7.5cm]{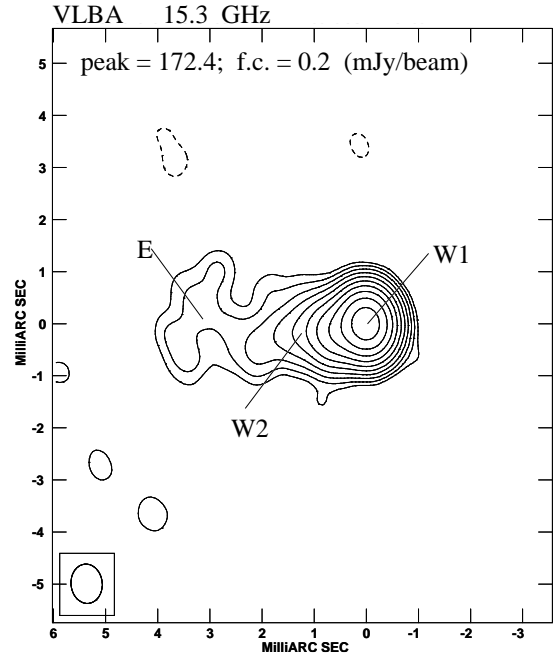}
\caption{VLBA image at 15.3 GHz of SBS 0846+513. Data are from the MOJAVE 
programme. On the image we provide the
peak flux density, in mJy/beam, and the first contour intensity (f.c.,
in mJy/beam) that corresponds to three times the noise measured on the
image plane. Contour levels increase by a factor of 2. The beam is
plotted on the bottom left corner of the image.}
\label{Fig10}
\end{figure}

\begin{figure}
\centering
\includegraphics[width=7.5cm]{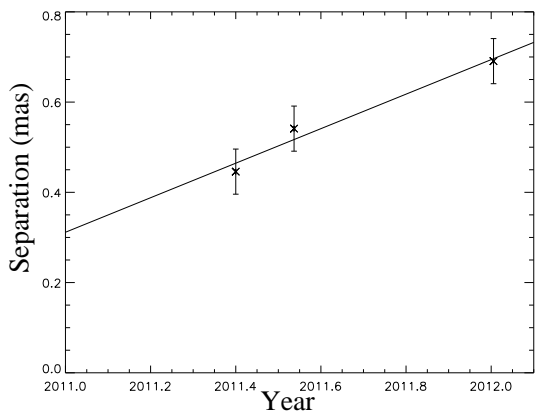}
\caption{Changes in separation with time between components W1 and W2. The
  solid line represents the regression fit to the 15-GHz MOJAVE data.}
\label{Fig11}
\end{figure}

The resolution of the VLA is not adequate to resolve
the radio structure of SBS\,0846+513 (Fig.~\ref{Fig7}) even when the array is
in its most extended 
configuration (full width half maximum $\leq$0.3 arcsec). When imaged with the high spatial resolution of the VLBA the source is
resolved in two components with a core-jet structure (Figs.~\ref{Fig8} and ~\ref{Fig9}), 
as also pointed out in a previous work by \citet{taylor05}. The core (component W) is still
unresolved with an upper limit of 0.3 mas, while the jet structure (component E) is 3.5
$\times$ 1.6 mas.
The flux density ratio between components E and W is $\sim$19 and 14 at 5 and 8.4 GHz
respectively. Their separation is about 3.5 mas, corresponding to $\sim$23 pc
projected distance, given the source's redshift. At 15 GHz component E shows an extended low-surface
brightness structure with a steep spectrum. On the other hand
component W is resolved into two compact components (labelled W1 and
W2 in Fig.~\ref{Fig10}). To investigate a possible proper motion of the jet, 
we compared the separation between W1, considered the core region, and
W2, assumed to be a knot in the jet, at the three observing epochs. To
this purpose, in addition to the data analysis on the image plane,
we model-fitted the visibility data using gaussian components of the
three-epoch MOJAVE data by means of the model-fitting option in \textsc{difmap}. This
approach is preferable in case we want to derive small variations in the
source structure and little changes in the position of the source components. From this comparison
we found that W1 and W2 are separating with an apparent velocity of
(0.38$\pm$0.07) mas/yr (Fig.~\ref{Fig11}), which corresponds to (8.2$\pm$1.5)$c$. This apparent
superluminal velocity suggests the presence of boosting effect. However, the availability of only three observing epochs spanning a
short time baseline without frequent time sampling implies large
uncertainties on the estimated values.\\
When we compare the total flux density derived from VLBA data with
that from the VLA, we find that the VLBA can recover at most 85\% of the VLA flux density, even when the
observations were performed simultaneously (i.e. 1996 January 6).
This may be related to the slightly different observing frequencies, 
i.e. 4.8 GHz at the VLA, whereas it is 5.0 GHz at the VLBA. 
However, in this case the 
spectral index would have to be extremely steep ($\alpha_r$ $\sim$4; $f_{\nu} \propto \nu^{-\alpha_{r}}$) to obtain such a
difference in the flux density, making this explanation unlikely.
Another possibility is related to observational limitations due to the
lack of short spacings of the VLBA array, implying that 
only structures smaller than $\sim$40 and 30 mas at 5 and 8.4 GHz, can be detected.
This suggests that the missing flux on the parsec scale may be related to 
extended, low-surface brightness features like a jet component resolved out by the VLBA array. \\

\begin{table*}
\caption{Log of the VLBA observations analysed in this paper and total
flux density. $^{a}$: data from the MOJAVE programme.}\label{vlba}
\begin{center}
\begin{tabular}{cccccc}
\hline \hline
Freq&Date&Code&Beam&Obs time&Flux\\
GHz &    &    &mas$\times$mas&min&mJy\\
\hline
 5.0&1995-03-15&BW015&3.99$\times$1.34&50&284$\pm$20\\
 5.0&1996-01-06&BW021&4.13$\times$1.29&50&281$\pm$20\\
 8.4&2011-03-12&BC196J&2.00$\times$0.97&10&210$\pm$16\\
15.3&2011-05-26&BL149DI$^{a}$&0.77$\times$0.59&30& 244$\pm$17\\
15.3&2011-07-15&BL149DM$^{a}$&0.76$\times$0.60&30& 213$\pm$15\\
15.3&2012-01-02&BL178AE$^{a}$&0.74$\times$0.64&36& 231$\pm$21\\
\hline
\end{tabular}
\end{center}
\end{table*}

\begin{table}
\caption{Flux density in mJy of the components of SBS\,0846+513 from VLBA
  data. Epochs of the observations are reported in Table~\ref{vlba}.} \label{vlba_results}
\begin{center}
\begin{tabular}{cccccc}
\hline \hline
 & E & W & W1 & W2\\
\hline
S$_{\rm 5 GHz, ep 1}$& 14$\pm$2 & 269$\pm$19 & - & - \\
S$_{\rm 5 GHz, ep 2}$& 17$\pm$2 & 265$\pm$18 & - & - \\
S$_{\rm 8.4 GHz}$& 14$\pm$2 & 196$\pm$14 & - & - \\
S$_{\rm 15.3 GHz, ep 1}$& 3.4$\pm$0.4 & 192$\pm$15 & 140$\pm$10 & 52$\pm$4 \\
S$_{\rm 15.3 GHz, ep 2}$& 3.1$\pm$0.3 & 240$\pm$24 &  175$\pm$17 & 65$\pm$6 \\
S$_{\rm 15.3 GHz, ep 3}$& 3.6$\pm$0.4 & 209$\pm$20 &  151$\pm$15 & 58$\pm$5 \\
\hline
\end{tabular}
\end{center}
\end{table}

\section{SED modeling}
\label{SED_modeling}

The $\gamma$-ray detection of the radio loud NLS1s has given us 
the possibility of studying the characteristics of this class of 
objects by modeling their broad band spectra.

We have built a non-simultaneous SED, all of which is based on
non-flaring data.  This ought to adequately represent the emission
from this object in a low state. The LAT spectrum was built with data
centred on 2010 October 4 to 2011 June 4 (MJD 55473--55716). In
addition we included in the SED the Effelsberg radio data collected on
2011 April 30 and the {\em Swift} (UVOT and XRT) data collected on
2011 September 15, thus well before and after the $\gamma$-ray
flare. The data from 2MASS All-sky PSC source catalogue and WISE
preliminary release Source Catalogue provided information about the IR
part of the spectrum. The flare centred on $\sim$\ MJD 55750 had a
variability timescale of $\sim 1$\ day, which constrains the size of
the emitting region during the flare.

\begin{figure} 
\centering
\vspace{2.0mm}
\includegraphics[width=7.0cm]{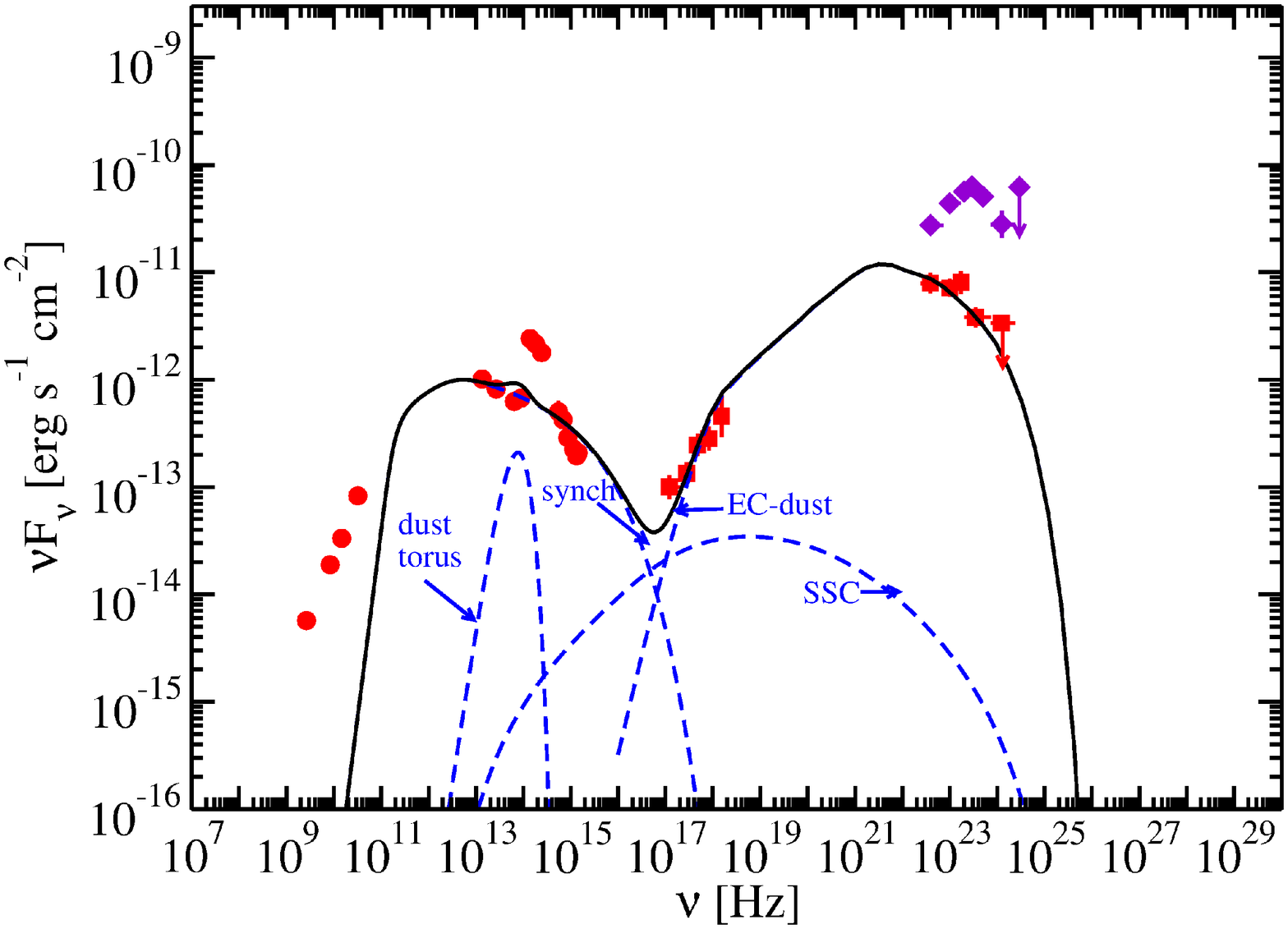}
\caption{Spectral energy distribution data (circles and squares) and model fit
  (solid curve) of SBS 0846$+$513 with the model components shown as dashed
  curves. The data points were collected by Effelsberg (2011 April 30), {\em
    Swift} (UVOT and XRT; 2011 September 15), and {\em Fermi}-LAT (2010 October 4--2011 June 4), together
  with archival data from 2MASS and WISE. The LAT spectrum during the flaring
  period (2011 June 4--July 4) is shown as the diamonds.}
\label{0846sed}
\end{figure}

We modeled the SED assuming emission from a relativistic jet with
mechanisms of synchrotron, synchrotron self-Compton (SSC), and
Compton-scattering of a dust torus external to the jet (EC-dust).  The
description of the model can be found in \citet{finke08_ssc} and
\citet{dermer09}. Jet powers were calculated assuming a two-sided jet. Although the flaring timescale does not constrain the emitting region during the low state, we have produced a model fit
which is roughly consistent with this timescale (about 1 day). The
synchrotron component can adequately explain the IR through UV points,
although some of the 2MASS points are significantly higher than the
rest, and are not well-fit.  This is probably due to contamination by
IR emission from the host galaxy. We note that this IR excess could be
consistent with the typical starlight expected from an elliptical
galaxy.  The radio spectrum is flat in $F_{\nu}$ (Fig.~\ref{Fig5and6}) and probably the result of a superposition of several
jet components \citep[e.g.,][]{konigl81}, so for our purposes these
points are considered upper limits. The synchrotron component we use
is self-absorbed below $\sim 10^{12}$\ Hz.  An attempt was made to fit
the X-ray through $\gamma$ ray data with an SSC component, but this
was not found to be possible.  Instead, an external component was
required. Correlations of $\gamma$-ray and optical flares with radio
light curves and rotations of optical polarization angles in
low-synchrotron-peaked blazars seem to indicate the
$\gamma$-ray/optical emitting region is outside the broad line region,
where the dust torus is the likely seed photon source
\citep[e.g.,][]{marscher10}. So we chose as our seed photon source a
dust torus, which was simulated as a one-dimensional ring with radius
$R_{dust}$ aligned orthogonal to the jet, emitting as a blackbody with
temperature $T_{dust}$ and luminosity $L_{dust}$. Observational
evidence for a dust torus in NLS1s as well as in the other Seyfert
galaxies was recently reported in \citet{mor12}. The fit can be seen
in Fig.~\ref{0846sed} and the parameters can be found in Table~\ref{table_fit}.  A description of the parameters can be found in
\citet{dermer09}. The dust parameters were chosen so that $R_{dust}$
is roughly consistent with the sublimation radius \citep{nenkova08},
assuming the torus luminosity is about 1/10 of the disc luminosity. If
the emitting region takes up the entire width of the jet, and the jet
is assumed to be conical, its half-opening angle would be about
1$^{\circ}$, approximately consistent with those measured from VLBI
observations of blazars \citep{jorstad05}. This size scale is roughly
the same as for the 15 GHz core (Fig.~\ref{Fig10}), and so if this
model fit is correct, the $\gamma$-ray emitting region is probably
located near this region, which is presumably a synchrotron
self-absorption photosphere.

The electron distribution used, a broken power-law with index $p_1=2.2$ below
the break at $\gamma^{\prime}_{brk}$ and $p_2=3.2$ for $\gamma^{\prime}_{brk}< \gamma^{\prime}$, is consistent with particles injected with index $2.2$, and emission
taking place in the {\em slow cooling regime} \citep[e.g.,][]{boett02}. That is, particles are
injected between $\gamma^{\prime}_{min}$ and $\gamma^{\prime}_{max}$, with a
cooling break at
$$
\gamma_{brk}\equiv \frac{3 m_ec^2}{4 c \sigma_T t_{esc} u_{tot} }\ 
$$ where $u_{tot}$ is the total energy density in the blob frame,
which in the case of our model fit is dominated by the external energy
density. The escape time $t_{esc}\approx R_b^\prime/c$ so that for our
model fit $\gamma^{\prime}_{brk}\sim 300$. Also note that in this
model fit the magnetic field and electrons are nearly in
equipartition.

Using the SDSS Data Release (DR) 1, \citet{zhou05} estimated the black
hole mass from several methods, the H$\beta$ broad line, the
[OIII]$\lambda$5007 narrow line, and the host galaxy's bulge
luminosity, as $\simeq$ $8.2\times10^6\ M_\odot$, $\simeq$ $5.2\times10^7\ M_\odot$, and
$\simeq$ $4.3\times10^7\ M_\odot$, respectively. Large uncertainties (0.4--0.7
dex) are associated to these techniques \citep{vestergaard04}. \citet{shen11} estimated the
BH mass based on the H$\beta$ broad line and MgII line in the SDSS DR7
spectrum as $\log(M_{BH}/M_\odot)$ = 7.99 $\pm$ 0.12 and
$\log(M_{BH}/M_\odot)$ = 7.79 $\pm$ 0.16, respectively. 
Note that the broadband SED shows no evidence for a blue bump, so the optical spectrum probably has
considerable contamination from jet emission not taken into account in \citet{shen11}, and so these mass
estimates should be taken with caution. If the \citet{zhou05} H$\beta$
estimate is correct, the total jet power will exceed the Eddington
luminosity for this source (~$L_{Edd} = 1.3\times10^{45}$ erg s$^{-1}\
M_{BH}/(10^7$\ M$_\odot$)~). Thus the BH mass is probably more on the
high side of these estimates, if the jet power is Eddington-limited.

The Compton dominance for SBS~0846+513, i.e., the ratio of the peak
luminosities of the Compton and synchrotron components, is $\approx
7$, which is a rather standard value for FSRQs, although quite a high
value for BL~Lacs \citep[e.g.,][]{giommi11,finke12}.  Its SED in many
ways resembles that of a FSRQ, with an X-ray spectral index $\sim
1.5$, consistent with values for FSRQs in the BAT \citep{ajello09} and
BeppoSAX \citep{donato05} catalogues.  As we discuss in section
\ref{Discussion}, its LAT flux and spectral index are also consistent
with values for FSRQs.

The flaring LAT spectrum is also shown in Fig.~\ref{0846sed}.
Unfortunately, there are no multiwavelength data simultaneous to this
flare.  For FSRQs, $\gamma$-ray flares are often associated with optical
flares, although there are occasions when they are not. Both types of
flares are seen, for example, from the FSRQ PKS~1510$-$089 \citep{marscher10} so
we can only guess at the behavior of SBS~0846+513 during this flare. However,
note that during the flare, the object reaches a luminosity of
$L_{\gamma,iso}\sim 1.0\times10^{48}$ erg s$^{-1}$, making the radiative power
$L_{\gamma,rad}\sim L_{\gamma,iso}/\Gamma^2\sim 4.4\times10^{45}$ erg s$^{-1}$,
assuming $\Gamma=15$. This is about 1/2 of the Eddington luminosity for a
$10^7$ M$_{\odot}$ black hole.

\begin{table*}
\footnotesize
\begin{center}
\caption{Model parameters for the SED shown in Fig.~\ref{0846sed}.\label{table_fit}}
\begin{tabular}{lcc}
\hline
Redshift & 	$z$		& 0.5835	  \\
Bulk Lorentz Factor & $\Gamma$	& 15	  \\
Doppler Factor & $\delta_{\rm D}$       & 15    \\
Magnetic Field & $B$         & 1.0 G   \\
Variability Timescale & $t_v$       & 1$\times$$10^5$ s \\
Comoving radius of blob & R$^{\prime}_b$ & 2.8$\times$10$^{16}$ cm \\
Jet Height & $r$ & $1.6\times10^{18}$\ cm \\
Low-Energy Electron Spectral Index & $p_1$       & 2.2     \\
High-Energy Electron Spectral Index  & $p_2$       & 3.2	 \\
Minimum Electron Lorentz Factor & $\gamma^{\prime}_{min}$  & $5.0$ \\
Break Electron Lorentz Factor & $\gamma^{\prime}_{brk}$ & $3.0\times10^2$ \\
Maximum Electron Lorentz Factor & $\gamma^{\prime}_{max}$  & $9.0\times10^3$ \\
Dust Torus luminosity & $L_{dust}$ & $4.0\times10^{44}$\ erg s$^{-1}$ \\
Dust Torus temperature & $T_{dust}$ & $1.5\times10^3$\ K \\
Dust Torus radius & $R_{dust}$ & $2.0\times10^{18}$\ cm \\
Jet Power in Magnetic Field & $P_{j,B}$ & $1.4\times10^{45}$\ erg s$^{-1}$ \\
Jet Power in Electrons & $P_{j,par}$ & $4.3\times10^{44}$\ erg s$^{-1}$ \\
\hline
\end{tabular}
\end{center}
\end{table*}

\section{Discussion and concluding remarks}
\label{Discussion}

After the 4 objects detected by {\em Fermi}-LAT in the first year of
operation \citep{abdo09c}, SBS 0846$+$513 is a new NLS1 detected by
{\em Fermi}-LAT during high $\gamma$-ray activity in 2011 June
\citep{donato11}. The power released by this object during the flaring
activity was a strong indication of the presence of a relativistic jet
as powerful as those in blazars, supported by the apparent
superluminal velocity of the jet derived by tracking the position of a
jet component in 2011-2012 VLBA data (see Sec.~\ref{radiomorph}).
Before the $\gamma$-ray flaring episode, the simultaneous
multifrequency observations performed by Effelsberg showed a flat
radio spectrum up to 32 GHz. After the flare, the spectral shape
changed, becoming convex. The spectral variability was also accompanied
by variations in the radio flux density, which were originally
detected at the higher frequencies, later moving to lower frequencies,
likely due to opacity effects. These spectral and variability
properties indicate blazar-like behaviour, which has already been
observed in other $\gamma$-ray NLS1s \citep{fuhrmann11}.

\begin{figure}
\includegraphics[width=7.5cm]{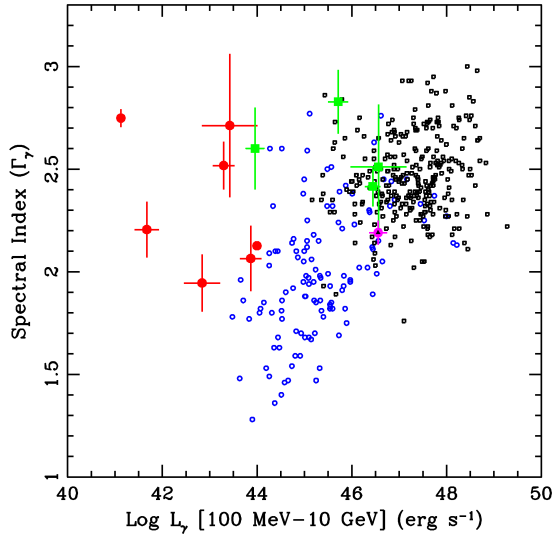}
\caption{The photon index $\Gamma_{\gamma}$ of FRI radio
  galaxies (red circles), FRII radio galaxies (green squares), BL Lacs
  (open blue circles) and FSRQs (open black squares) are plotted
  together with the NLS1 SBS 0846$+$513 (magenta point including a
  black triangle) as a function of the observed isotropic $\gamma$-ray
  luminosity (100 MeV--10 GeV). Adapted from \citet{abdo10b}.}
\label{Lgamma}
\end{figure}

\citet{ghisellini08, ghisellini11}, investigating the $\gamma$-ray
properties of blazars detected by {\em Fermi}-LAT in the first year of
observation, suggested a transition between BL Lac objects and FSRQs
that can be justified mainly by the different accretion regimes:
highly sub-Eddington in the former case, near-Eddington in the
latter. In this context the high accretion rate of SBS 0846$+$513, and
more generally of the $\gamma$-ray radio-loud NLS1s, is another
indication of its similarity with FSRQs.  A comparison of the BL Lacs
and FSRQs from the First LAT AGN Catalog \citep[1LAC;][]{abdo1lac}
with ``misaligned AGN'' detected by {\em Fermi} (MAGN; non-blazar AGN
with jets pointed away from the observer) in the $\Gamma_{\gamma}$ -
$L_{\gamma}$ plane has shown that MAGN and blazars occupy different
regions of the plane, with only two high redshift FRII galaxies, 3C
207 and 3C 380, which lie among the FSRQs \citep{abdo10b}. This is in
agreement with the idea that MAGN are less beamed than blazars.  In
this context it is interesting to consider also the $\gamma$-ray
NLS1s. For a direct comparison with the results shown in
\citet{abdo10b} we calculated the flux and photon index over the third
year of {\em Fermi} operation in the 0.1-10 GeV energy band, resulting in
$\Gamma_{\gamma}$ = 2.19 $\pm$ 0.06 and Flux (0.1--10 GeV) =
(6.6$\pm$0.6) $\times$ 10$^{-8}$ erg cm$^{-2}$ s$^{-1}$. The
corresponding observed isotropic $\gamma$-ray luminosity in the 0.1--10
GeV is 3.6$\times$10$^{46}$ erg s$^{-1}$. We plotted these values of
SBS 0846$+$513 in the $\Gamma_{\gamma}$ - $L_{\gamma}$ plane together
with the FSRQs, BL Lacs, and misaligned AGN from \citet{abdo10b}. As can be seen SBS 0846$+$513 lies in the blazar-region,
in particular in the transition region between the distribution of BL
Lacs and FSRQs (Fig.~\ref{Lgamma}). We note that also the other
$\gamma$-ray NLS1 observed in flaring activity, PMN J0948$+$0022
\citep[$\Gamma_{\gamma}$ = 2.26 $\pm$ 0.08 and 0.1--10 GeV luminosity of
9.6$\times$10$^{46}$ erg s$^{-1}$ over the first 24-month of {\em Fermi}
operation;][]{grandi11}, occupies the same blazar-region in that plane. This should reflect a similar viewing angle with respect to the
jet axis and beaming factors for the $\gamma$-ray emission between blazars and
the two $\gamma$-ray NLS1s SBS 0846$+$513 and PMN J0948$+$0022. In the same way the spectral
evolution during the flaring activity in June 2011 observed in $\gamma$ rays from SBS 0846$+$513 is a common behaviour in bright
FSRQs and low-synchrotron-peaked BL Lacs detected by {\em Fermi} \citep{abdo10d}, with a change in photon index $<$ 0.2--0.3 and an
increasing spectral curvature.

One of the key questions is the maximum power released by the jets of
NLS1s. During the $\gamma$-ray flaring activity observed in 2011 June--July SBS 0846$+$513 reached an observed isotropic $\gamma$-ray
luminosity (0.1--300 GeV) of 1.0$\times$10$^{48}$ erg s$^{-1}$ on
daily timescales, comparable to that of luminous FSRQs. After PMN
J0948$+$0022 \citep{foschini_etal11}, this is the second NLS1 observed
to generate such a high power. This could be an indication that all
the radio-loud NLS1s are able to host relativistic jets as powerful as
those in blazars, despite the lower BH mass; alternatively some NLS1s
could have peculiar characteristics allowing the development of these
relativistic jets. The mechanism at work for producing a relativistic
jet is not clear, and the physical parameters that drive jet formation
is still under debate. One fundamental parameter could be the black
hole mass, with only large masses allowing for the efficient formation of a relativistic jet. It was 
for example noted by some authors \citep{mclure04, liu06, sikora07} that quasars with $M_{\rm BH} >
10^{8}\ M_\odot$ reach radio loudness 3 orders of magnitude greater
than quasars with $M_{\rm BH} < 3 \times 10^{7}\ M_\odot$. The large
  radio loudness of SBS 0846$+$513 could challenge this idea if the black hole
  masses estimated by \citet{zhou05} are confirmed. According
to the ``modified spin paradigm'' discussed in \citet{sikora07}, another fundamental
parameter for the efficiency of a 
relativistic jet production should be the BH spin, with SMBHs in elliptical galaxies having on average
much larger spins than SMBHs in spiral galaxies. This
is due to the fact that spiral galaxies are characterized by multiple
accretion events with random orientation of angular momentum vectors
and small increments of mass, while elliptical galaxies underwent at
least one major merger with large matter accretion triggering an
efficient spin up of the SMBH.  The accretion rate (thus the mass) and
the spin of the black hole may be related to the host galaxy, leading
to the hypothesis that relativistic jets can develop only in
elliptical galaxy \citep[e.g.][]{marscher09}.

In this context the discovery of relativistic jets in a class of AGN
usually hosted by spiral galaxies, such as the NLS1s, was a great
surprise.  Unfortunately only very sparse observations of the host
galaxy of radio-loud NLS1s are available at this time, and the sample
of objects studied by \citet{deo06} and \citet{zhou06} had $z < 0.03$
and $z < 0.1$, respectively, including both radio-quiet and radio-loud
objects. We note that BH masses of radio-loud NLS1s are generally
larger with respect to the entire sample of NLS1s \citep[$M_{\rm BH}
\approx (2-10) \times 10^{7} M_\odot$][]{komossa06}, even if still
small when compared to radio-loud quasars. The larger BH masses of
radio-loud NLS1s with respect to radio-quiet NLS1s could be related to prolonged accretion episodes that
can spin-up the BHs. The small fraction of radio-loud NLS1s with
respect to radio-loud quasars could be an indication that in the
former the high accretion usually does not last sufficiently long to
significantly spin-up the BH \citep{sikora09}.

Of the $\gamma$-ray emitting NLS1s, host galaxy imaging is available
only for 1H 0323$+$342, with HST and the Nordic Optical
Telescope. These observations reveal a one-armed galaxy morphology or
a circumnuclear ring, suggesting two possibilities: the spiral arms of
the host galaxy \citep{zhou07} or the residual of a merging galaxy
\citep{anton08}. No significant resolved structures have been observed
by HST for SBS 0846$+$513 \citep{maoz93}, and no high-resolution
observations are available for the remaining $\gamma$-ray NLS1s.  Thus
the possibility that the development of relativistic jets in these
objects could be due to strong merger activity, is not ruled
out. Further high-resolution observations of the host galaxies of
$\gamma$-ray NLS1s will be fundamental to obtain important insights
into relativistic jet formation and development.

To conclude, SBS 0846$+$513 shows all the characteristics of the
blazar phenomenon. The extreme power released by SBS 0846$+$513 during
the high $\gamma$-ray activity in 2011 July confirms that, as with PMN
J0948$+$0022, NLS1s can host relativistic jets as powerful as
blazars. Radio and $\gamma$-ray data collected for SBS 0846$+$513
suggest spectral and variability properties similar to blazars, and the
modeling of the average SED gives similar results to those of blazars,
including similar Lorentz factors. This could be an indication that
these $\gamma$-ray NLS1s are low mass (and possibly younger) version
of blazars. The detection of new radio-loud NLS1s in $\gamma$ rays by
{\em Fermi}-LAT will be important for extending the source sample, and
better characterizing this new class of $\gamma$-ray emitting
AGN. Equally important will be to perform further multifrequency
observations of the $\gamma$-ray emitting NLS1s already detected by
{\em Fermi} and investigate in detail their characteristics over the
entire electromagnetic spectrum, to help understand their nature.

\section*{Acknowledgments}

The {\em Fermi} LAT Collaboration acknowledges generous ongoing
support from a number of agencies and institutes that have supported
both the development and the operation of the LAT as well as
scientific data analysis.  These include the National Aeronautics and
Space Administration and the Department of Energy in the United
States, the Commissariat \`a l'Energie Atomique and the Centre
National de la Recherche Scientifique / Institut National de Physique
Nucl\'eaire et de Physique des Particules in France, the Agenzia
Spaziale Italiana and the Istituto Nazionale di Fisica Nucleare in
Italy, the Ministry of Education, Culture, Sports, Science and
Technology (MEXT), High Energy Accelerator Research Organization (KEK)
and Japan Aerospace Exploration Agency (JAXA) in Japan, and the
K.~A.~Wallenberg Foundation, the Swedish Research Council and the
Swedish National Space Board in Sweden. Additional support for science
analysis during the operations phase is gratefully acknowledged from
the Istituto Nazionale di Astrofisica in Italy and the Centre National
d'\'Etudes Spatiales in France.

We thank the Swift team for making these observations possible, the
duty scientists, and science planners. This research has made use of
data from the MOJAVE database that is maintained by the MOJAVE team
(Lister et al. 2009, AJ, 137, 3718). The OVRO 40-m monitoring program
is supported in part by NASA grants NNX08AW31G and NNX11A043G, and NSF grants AST-0808050 
and AST-1109911. This paper is partly based on observations with the
100-m telescope of the MPIfR (Max-Planck-Institut f\"ur
Radioastronomie) at Effelsberg and the Medicina telescope operated by
INAF--Istituto di Radioastronomia. We acknowledge A. Orlati,
S. Righini, and the Enhanced Single-dish Control System (ESCS)
Development Team. We acknowledge financial contribution from agreement
ASI-INAF I/009/10/0.  This publication makes use of data products from
the Two Micron All Sky Survey, which is a joint project of the
University of Massachusetts and the Infrared Processing and Analysis
Center/California Institute of Technology, funded by the National
Aeronautics and Space Administration and the National Science
Foundation.  This publication makes use of data products from the
Wide-field Infrared Survey Explorer, which is a joint project of the
University of California, Los Angeles, and the Jet Propulsion
Laboratory/California Institute of Technology, funded by the National
Aeronautics and Space Administration.  We thank the anonymous referee for useful suggestions. F. D'Ammando would like to thank
Gino Tosti and Marco Ajello for fruitful comments and discussions,  and Paola Grandi who has made the data of her paper available.

\end{document}